\documentclass[11pt]{article}
\pdfoutput=1 
\usepackage{graphicx}
\usepackage{dcolumn}
\usepackage{bm}
\usepackage{float}
\usepackage{placeins}
\usepackage{hyperref}
\usepackage{natbib}

\usepackage{subfiles}

\usepackage[T1]{fontenc}
\usepackage[utf8]{inputenc}
\usepackage{authblk}

\title{Imprints of Gravitational Millilensing on the Light Curve of Gamma-Ray Bursts}
\author[1]{Zeinab Kalantari\thanks{zeinab.kalantari@physics.sharif.edu}}
\author[2]{Alaa Ibrahim}
\author[1]{Mohammad Reza Rahimi Tabar}
\author[1]{Sohrab Rahvar}
\affil[1]{Department of Physics, Sharif University of Technology, Tehran 11365-9161, Iran}
\affil[2]{Department of Physics, College of Science, P.O.Box 36, P.C. 123, Muscat, Sultanate of Oman}

\begin{document}




\maketitle



\begin{abstract}
In this work, we search for signatures of gravitational millilensing in Gamma-ray bursts (GRB) in which the source-lens-observer geometry produces two images that manifest in the GRB light curve as superimposed peaks with identical temporal variability (or echoes), separated by the time delay between the two images. According to the sensitivity of our detection method, we consider millilensing events due to point mass lenses in the range of $10^5 - 10^7  M_{\odot}$ at lens redshift about half that of the GRB, with a time delay in the order of $10$ seconds. Current GRB observatories are capable of resolving and constraining this lensing scenario if the above conditions are met. We investigated the Fermi/GBM GRB archive from the year 2008 to 2020 using the autocorrelation technique and we found one millilensed GRB candidate out of 2137 GRBs searched, which we use to estimate the optical depth of millilensed GRBs by performing a Monte-Carlo simulation to find the efficiency of our detection method. Considering a point-mass model for the gravitational lens, where the lens is a supermassive black hole, we show that the density parameter of black holes ($\Omega_{BH}$) with mass $\approx10^6 M_\odot$ is about $0.007 \pm 0.004$. Our result is one order of magnitude larger compare to consist with previous work in a lower mass range ($10^2 - 10^3  M_{\odot}$), which gave a density parameter $\Omega_{BH} \approx 5\times 10^{-4}$, and recent work in the mass range of $10^2 - 10^7 M_{\odot}$ that reported $\Omega_{BH} \approx 4.6\times 10^{-4}$. The mass fraction of black holes in this mass range to the total mass of the universe would be $f\approx \Omega_{BH}/\Omega_M=0.027 \pm 0.016$.

\begin{description}
\item[Keywords]
Gamma Ray Bursts, Gravitational microlensing, Black holes
\end{description}
\end{abstract}
\maketitle



\section{INTRODUCTION}
\label{sec:Introduction}
Gamma-ray bursts (GRBs) are one of the most luminous and prolific high-energy transients that have been observed since their discovery in 1963 by the United States’ Vela satellites and \cite{Vela} wrote the first publication of GRBs when 16 GRBs were detected by Vela.
The understanding of the origin of GRBs started with the launch of the Burst and Transient Source Experiment (BATSE) onboard the Compton Gamma-ray Observatory (CGRO) in 1991, which detected through 1999 over 3000 gamma-ray bursts and established the isotropic distribution of GRBs in the sky (\cite{BATSE,BATSEIsotrop}).
Accurate and rapid localization of GRB hosts was made possible upon the launch of BeppoSAX in 1996, which enabled the search for burst afterglow at longer wavelengths (\cite{3a}).
Spectroscopy of the fading optical counterparts of GRBs with ground telescopes, starting 1997, allowed measurement of the redshift and pinpointing the host galaxies, which showed that GRBs originate beyond the Milky Way galaxy, hence confirming their cosmological origin (\cite{Metzger}). Our understanding of GRBs has been enhanced even further with the launch of the Swift and Fermi gamma-ray observatories in 2004 and 2008, respectively (\cite{6a,FermiInstroment}),  which helped reveal new populations of GRBs, including GRBs at high redshift ($z \sim8$) (\cite{6b}), ultra-long GRBs, and ultra-high energy (GeV) GRBs (\cite{6c}).
\par
Theoretical investigations of gravitational lensing of GRBs have been carried out for many years.  \cite{Paczynski86} proposed the idea that the GRBs could be gravitationally lensed with the observable signature of multiple images of the same GRB registered/received as light curves at different times due to the different light paths of the different images. \cite{Marani98} compared 1235 bursts from the BATSE catalog to find gravitational lensed GRBs by galaxies along the line of sight and reported null results. Previous works on searching for gravitationally lensed GRBs focused mostly on strong lensing where the time delay between the images is greater than the duration of the burst. Similarities among pairs of GRBs light curves with identical spectra and close locations in the sky as the primary search criteria were utilized in multiple searches (\cite{2sim,4sim,8sim,10sim,12sim}). Null results of these investigations did not rule out the presence of strong gravitationally lensed GRBs but reduced its likelihood and set upper limits to the lensing optical depth (\cite{10sim}). In particular, the uncertainties due to Poisson noise in the GRBs light curve (at low signal-to-noise ratios) may lead to notable differences between the gravitationally lensed images that are recorded as separate GRBs.
Furthermore, some gravitationally lensed images of GRB light curves could actually be millilensed, leading to time delays shorter than the burst duration, and hence separate pairs of similar light curves would not be resolvable. Instead, images would appear blended in the same GRB light curves but can manifest as local peaks separated by the time delay, with the peaks (or echos) corresponding to the unresolved lensed images (\cite{3sim,Population3,Ji2018StrongLO}). 
\par
\cite{6sim} studied the time delay probability distribution of GRB lensed by point masses and found it to be a broad function that peaks at $\Delta t= 50 s\ M_L/10^6M_{\odot}$ then rapidly drops as the time delay increases. \cite{3sim} searched the entire BATSE catalog (1991-2000) to find these types of short echoes (with a time delay of several seconds, comparable to the burst duration) and found 11 probable gravitationally lensed GRB candidates (out of 1512 GRBs) for the lens model of a globular cluster with mass of $\approx 10^5 -10^7 M_{\odot}$. \cite{Population3} applied the autocorrelation technique to GRB light curves in the BATSE catalog and found one candidate (GRB 940919) of a lensed gamma-ray burst by a Population $III$ star out of 1821 GRB searched, with a lens mass range of $10^ 2 -10^ 3 M_{\odot}$ and redshift range of 20 to 200, which induced a time delay about 1 second between two GRB images. \cite{Ji2018StrongLO} searched for echoes in GRB light curves due to strong lensing by massive compact dark matter halo with mass greater than the $10 M_{\odot}$ and a time delay of $\geq 1$ ms using the Fermi/GBM and Swift/BAT archives but did not find any lensed candidates. Recently, \cite{2021NatAs} used Bayesian analysis to identify gravitational lensing in the mass range of  $10^2-10^7 M_{\odot}$ in $\approx 2700$ short and long GRBs of BATSE dataset and found one candidate.
\par
Given that GRBs are at cosmological distances, gravitational lensing can provide an opportunity to estimate the number density of massive compact objects by calculating the rate of GRB lensing. \cite{7sim} searched for lensing effects in the first 44 BATSE GRBs and used the absence of any gravitationally lensed candidate to exclude compact objects with $10^6-10^8 M_{\odot}$ from the density of compact objects in the universe. \cite{marani1999} (1999) investigated the assumption that dark matter compact objects with density parameter $\Omega=0.15$, dominated by black holes of mass $\approx 10^{6.5} M_{\odot}$ make a considerable fraction of the mass in the universe and ruled out this model due to the null result of searching for gravitationally lensed GRBs with BATSE and Ulysses data.
Recently, \cite{Oguri} estimated the expected observed rates of strongly lensed GRBs with redshift between 0.5 and 4 and found the rate to grow with redshift from $0.01$ (at $z=0.5$) to 1 (at $z=4$), in units of lensed GRB per year per the entire sky, with long and short GRBs giving similar rates at the same redshift. 
\par
In this work, we search for millilensed GRB candidates using the Fermi Gamma-Ray Burst Monitor (GBM), which records GRBs in a broad energy range (8 keV to 40 MeV) with a field of view of $\geq 8$ sr (\cite{4th,FermiInstroment}). We present the results of GRBs recorded between 2008 and 2020 (\cite{DataCatal}) and focus on the lensing effect that gives rise to two images that manifest in the light curve as local peaks with identical temporal variability (echoes), but different magnification (count rate), separated by the time delay between the two images.
We model compact objects with point mass and assume a uniform distribution of these compact objects in the comoving volume of the Universe. Due to the sensitivity of our detection method, we focus on searching  for gravitational lenses with mass in the order of million solar mass that produce gravitational images of a GRB with angular separation in the order of 0.01 arc-second which is usually termed as millilensing (\cite{millilensing}). We use the autocorrelation technique to find the time delay between the echo signals that have identical temporal variability within the same GRB light curve. We just consider long GRBs, which are more relevant to the time delay considered. We apply a Monte-Carlo simulation to produce mock millilensed GRBs light curves to find the efficiency and accuracy of the autocorrelation technique.
\par
We used data from the GBM instrument of the Fermi telescope to obtain the light curves of GRBs. Fermi/GBM has twelve thallium-activated sodium iodide (NaI) and two bismuth germanate (BGO) detectors (\cite{FermiInstroment}). We extracted light curves from the NaI detectors with the highest signal using the time-tagged event (TTE) files and considered long GRBs (with $T_{90} \geq 2 s$) in our data analysis, which we probed for time delays less than 300 seconds. Our final goal is to estimate the contribution of heavy black holes in the matter content of the universe by GRB lensing analysis.
\par
The paper is organized as follows. Section \ref{sec:GLensing} presents a brief overview of the theory of gravitational lensing as it applies to GRBs. Section \ref{sec:Methodology} describes our method to estimate the mass of the gravitational lens by applying it to a simulated lensed GRB light curve and presents the efficiency and  precision of our method of finding millilensed candidates. In section \ref{sec:DataAnalysis} we report the result of our search to find probable candidates of lensed GRBs among those recorded by Fermi/GBM in 2008-2020. Section \ref{sec:Results} discusses the characteristics of the gravitationally lensed GRB candidate, including the mass range, the lensing probability, and an estimate of the number density for such compact objects in the universe.

\section{THEORY OF GRAVITATIONAL MILLILENSING}
\label{sec:GLensing}
When a massive object is located close to the line of sight between an observer and a source, gravitational lensing occurs. A point mass gravitational lens magnifies and makes two different images of the source. In this study, a GRB plays the role of the source, and each image with a gravitationally-induced time delay and different magnification can be detected through the observed burst light curve. 
The angular resolution of Fermi/GBM ($\sim 2^\circ$) is not sufficient to resolve two gravitationally lensed images with small angular separation. However, lensing events with closely displaced images and shorter time delays will result in detectable repeated structures or echoes at different magnifications in the same GRB light curve. This section is dedicated to a brief review of gravitational lensing formalism.
\subsection{Magnification Ratio and Time Delay}
\label{subsec:MagnificationDelay}
\begin{figure}[!htbp]
\centering
\includegraphics[width=\linewidth]{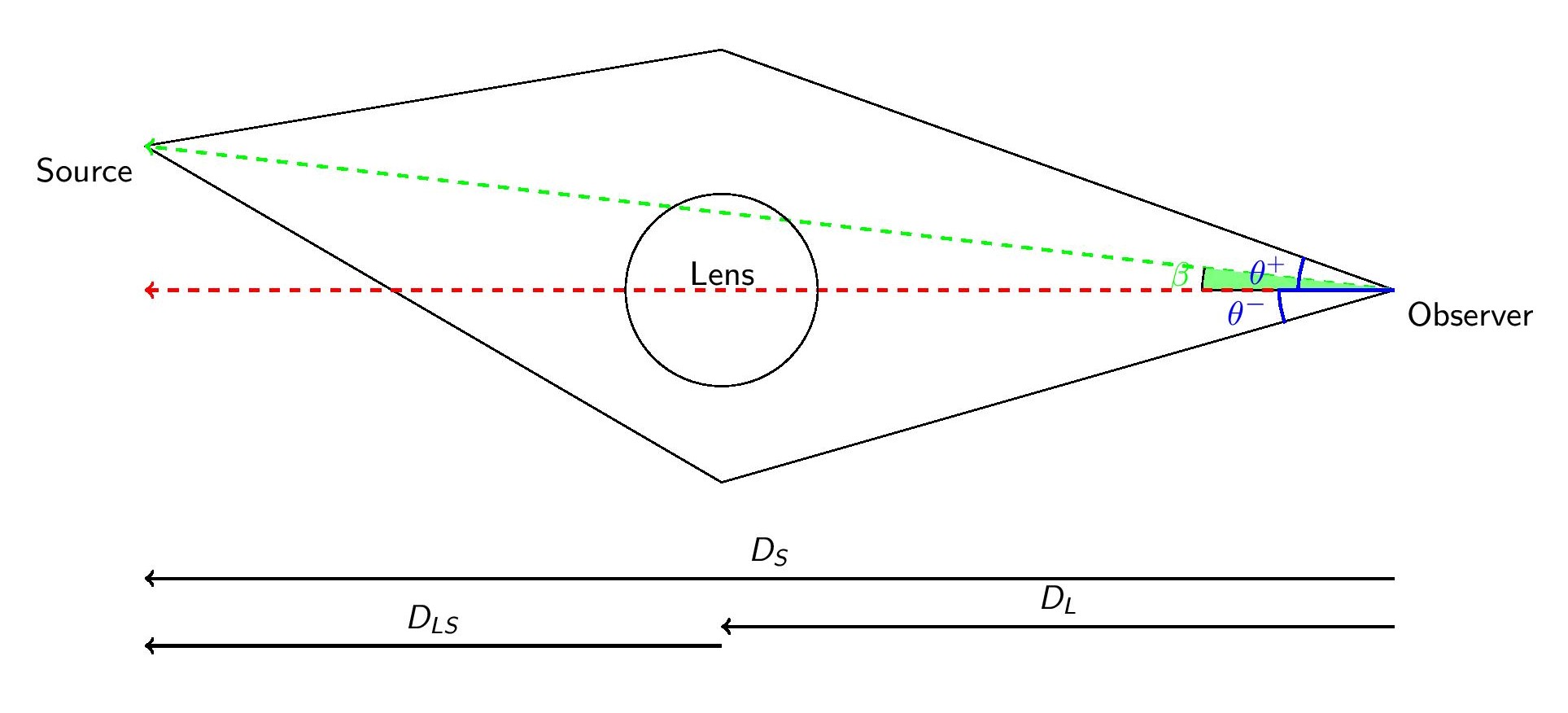}
\caption{Configuration of photons paths of two millilensed images of a source. According to Eq.(\ref{eq:Fermatdelay}), the millilensed negative parity image arrives later than the negative parity image to the observer.}
\label{fig:ConfigurationImages}
\end{figure}

Gravitational lensing produces two images from a source with the angular positions of
\begin{equation}
\theta_{\pm}=\frac{\beta \pm \sqrt{\beta^2 + 4 \theta _E ^2 }}{2}, 
\label{eq:theta}
\end{equation}
where $\beta$ is the position of the source in the absence of a lens, also called impact parameter and $\theta_E$ is the Einstein angle which is defined as
\begin{equation}
\theta_{E}=2\sqrt{ \frac{GM_L}{c^2} \frac{D_{LS}}{D_L D_s}},
\label{eq:EinsteinRadius}
\end{equation}
where $D_{S}$, $D_{L}$, and $D_{LS}$ are the angular diameter distances from the observer to the source and to the lens, and the angular diameter distance between lens to the source, respectively, as shown in Fig.\ref{fig:ConfigurationImages}
\par
Two parameters in gravitational lensing have a significant role in our analysis of the GRBs light curve. First, the time delay between the two images is based on the assumption of the point mass model for the lens mass. \cite{Narayan} derived the time delay from Fermat's potential by
\begin{equation}
\Psi(\theta)=\frac{1}{2 \theta_{E}^2}(\theta -\beta)^2 - \ln |\theta|.
\label{eq:fermat}
\end{equation}
Fermat's potential for the two images with positive and negative parities can be found by
\begin{equation}
\Psi(\theta_{\pm})=\frac{1}{2 \theta_{E}^2}(\theta_{\pm} -\beta)^2 - \ln |\theta_{\pm}|,
\label{eq:fermatPM}
\end{equation}
So the difference between Fermat's potential for the two images is
\begin{equation}
\Psi(\theta_{+})-\Psi(\theta_{-})=\frac{1}{2 \theta_{E}^2}[(\theta_{+} -\beta)^2 -(\theta_{-} -\beta)^2]- \ln | \frac{\theta_{+}}{ \theta_{-}}|,
\label{eq:DeltaFermat1}
\end{equation}
Substituting Eq.(\ref{eq:theta}) in Eq.(\ref{eq:DeltaFermat1}) equation leads to
\begin{equation}
\Psi(\theta_{+})-\Psi(\theta_{-})=\frac{-1}{2 \theta_{E}^2}[\theta_{+}^2- \theta_{-}^2 ]- \ln | \frac{\theta_{+}}{ \theta_{-}}|
\label{eq:DeltaFermat2}
\end{equation}
The time delay between the two images is proportional to the difference between the Fermat's potential of the images and is given by
\begin{equation}
t^+ - t^-= \frac{1+z_L}{c} \frac{D_{L}D_{S}}{D_{LS}} \theta_E^2[\Psi(\theta_{+})-\Psi(\theta_{-})]
\label{eq:Fermatdelay}
\end{equation}
By defining the normalized impact parameter as $y=\frac{\beta}{\theta_E}$ the time delay can be calculated as
\begin{equation}
\Delta t= 4\frac{GM_L}{c^3} (1+z_L)[\frac{y}{2}\sqrt{y^2+4}+\ln(\frac{\sqrt{y^2+4}+y}{\sqrt{y^2+4}-y})],
\label{eq:delay}
\end{equation}
where $M_L$ is the lens mass and $\Delta t = t^- - t^+$ is the difference in arrival time of light to the observer from the negative and positive parity images and $z_L$ is the redshift of the lens. $\Delta t$ is positive when $t^- \geq t^+$, so the negative parity image will arrive to the observer later than the positive parity image (Fig.\ref{fig:ConfigurationImages}). The second parameter is the flux ratio of the two images which is given by the ratio of the two magnification factors 
\begin{equation}
R=|\frac{\mu_+}{\mu_ -}|=\frac{y^2+2+y \sqrt{y^2+4}}{y^2+2-y \sqrt{y^2+4}},
\label{eq:magnification}
\end{equation}
where $\mu_{\pm}$ are the magnification factors of the positive and negative parity images. If we assume a lens mass of $5 \times 10^5
M_\odot$ and set the cosmological redshift of the lens at half of the GRB distance ($z_L=1\sim1Gpc$), the angular separation of two images $|\theta_+-\theta_-|$ would be in the order of $10^{-2}$ arcsecond. Therefore, two gravitationally lensed images of a GRB will appear as one GRB with the two images appearing as consecutive peaks or sub-bursts in the same light curve when the time delay between the two images is less than the recording time of the GRB observation and longer than the burst duration.

\subsection{Optical Depth}
\label{subsec:OpticalDepth}
\begin{figure}[!htbp]
\centering
\includegraphics[width=0.7\linewidth]{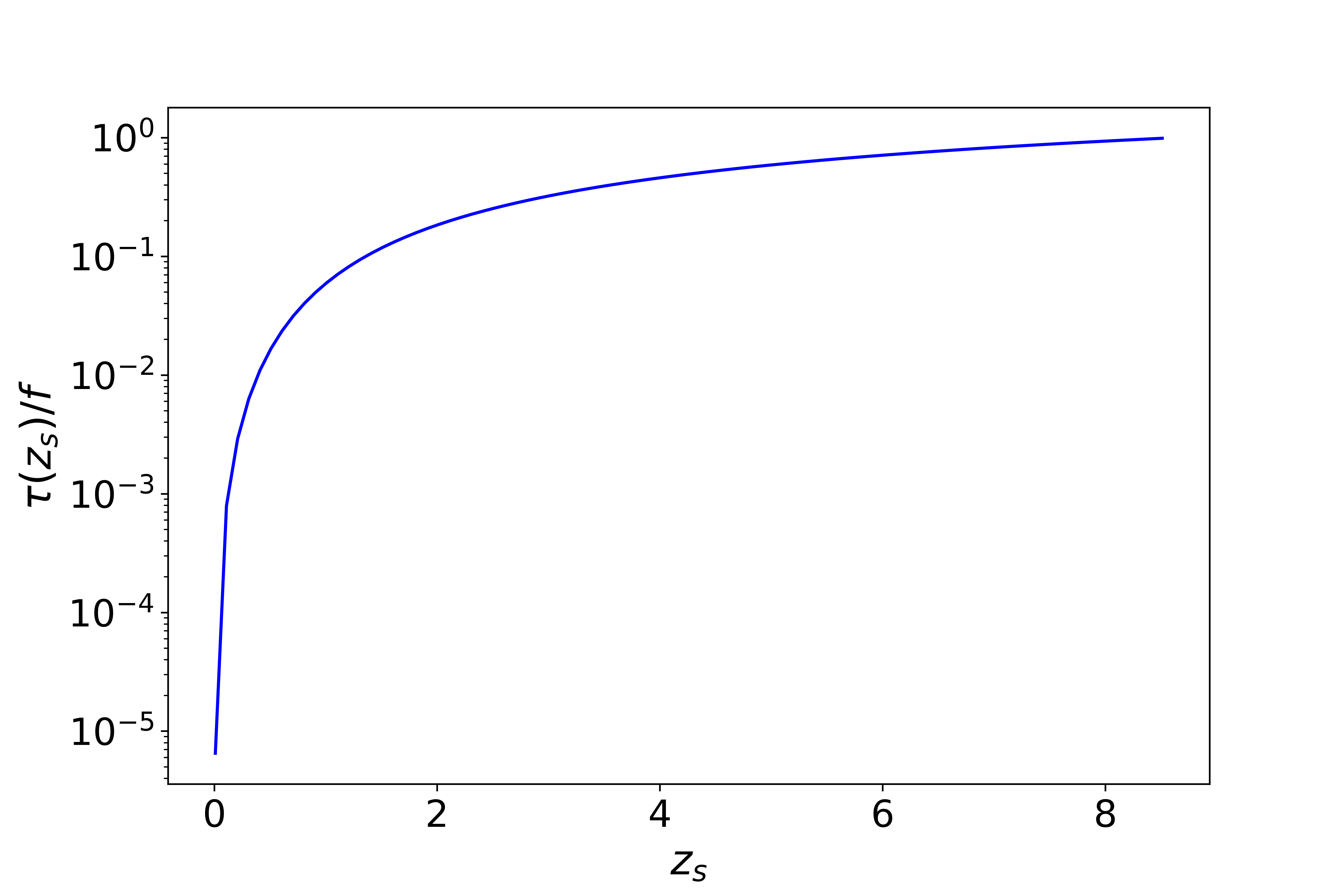}
\caption{Optical depths as a function of GRB redshift ($\tau (z_s)/f$)for a cosmological model with density parameters of $\Omega_\Lambda =0.74$ and $\Omega_M =0.26$.}
\label{fig:opticalDepth}
\end{figure}

In this section, we investigate the probability of GRB millilensing. Since GRBs are located at the higher redshifts, the gravitational millilensing of these events is highly probable.
For simplification, we assume that gravitational lenses are uniformly distributed in the comoving volume and assign $\rho_L$ as the comoving mass density of lenses. We define the fraction of the mass of compact objects to the total matter mass as
\begin{equation}
f=\frac{\rho_{L}}{ \rho_{M}}=\frac{\Omega_{L}}{ \Omega_{M}}
\label{eq:fDM}
\end{equation}
 where $0\leq f\leq 1$. Here $\rho_L$ is the overall density of compact lenses and $\rho_M$ is the density of matter ($\rho_M= \rho_{cr}\Omega_M$).The optical depth defines as the probability of lensing with a source at cosmological redshift ($z_s$) as \cite{Ji2018StrongLO} and given by
\begin{equation}
\tau (z_s) =\int _0 ^{z_s} \sigma \frac{\rho_L}{M_L}   (1+z_L)^2 \frac{ c\ dz_L}{H(z_L)},
\label{eq:tauS}
\end{equation}
where $\sigma$ is the effective lens size defined by Einstein cross-section 
\begin{equation}
\sigma=\pi R_E^2,
\label{eq:crossSection}
\end{equation}
and $R_E$ is the Einstein radius of the lens in the lens plane and is equal to
\begin{equation}
R_E= D_L\theta_E.
\label{eq:EinRadius}
\end{equation}
Since $R_E^2 \propto M_L$, then $\sigma \frac{ \rho_L}{M_L}$ would be proportional to the mass density of lenses. This means that the optical depth is independent of the mass function of lenses. Substitution of equations (\ref{eq:fDM}), (\ref{eq:crossSection}) and (\ref{eq:EinRadius}) into Eq.(\ref{eq:tauS}) results in
\begin{equation}
\tau (z_s) =f \frac{3 \Omega_M H_0}{2 c}\int _0 ^{z_s}  \frac{ (1+z_L)^2 D_{LS} D_L}{D_S \sqrt{\Omega_M (1+z_L)^3+\Omega_{\Lambda}}} dz_L,
\label{eq:tauSSimplified}
\end{equation}
For a cosmological model with Hubble parameter of $H=100 h \frac{km/s}{Mpc}$ where $h=0.7$ and density parameters of $\Omega_\Lambda =0.74$ and $\Omega_M =0.26$, $\tau (z_s)/f$ is depicted in Fig.\ref{fig:opticalDepth}. According to this figure, we anticipate that the gravitational lensing rate of GRBs will increase as we probe larger cosmological redshifts.
We use the distribution of the GRB redshifts presented in the second catalog of Fermi (\cite{2nd}) to find the average optical depth form
\begin{equation}
\tau_{Average} =\frac{\int_0^{z_{max}} \frac{dN(z)}{dz} \ \tau (z)\ d z}{\int_0^{z_{max}} \frac{dN(z)}{dz}  d z} =\frac{N_{Detected}}{N_{Total}},
\label{eq:tauTotal}
\end{equation}
where $N_{Detected}$ is the number of detected events with millilensing signature and $N_{Total}$ is the overall number of GRB events, where $\frac{dN(z)}{dz}$ is the differential number of GRBs at a given cosmological redshift of $z$. With the redshift distribution of the GRBs (\cite{2nd}) and based on standard cosmological model, we find $\tau_{Average}=0.217 f$. In section \ref{sec:Methodology},  $N_{Detected}$  is corrected by the efficiency of our detection method.

\section{METHODOLOGY}
\label{sec:Methodology}
While the sub-arcsecond angular separation of two images of a GRB lensed by the mass stated before will not be resolved by Fermi/LAT and the GRB will be recorded as one event, the light curve recorded by Fermi/GBM will be a superposition of the two light curves corresponding to the two lensed images and this can be resolved when the time delay is less than the GRB recording time and larger than burst duration. \cite{Ji2018StrongLO} used the autocorrelation function to probe this scenario and extract the time delay in the order of 1 ms in the candidate GRB light curves due to lenses with the mass range of $10^2 M_{\odot} \leq M_L\leq 10^3 M_{\odot}$ This approach is optimized for Fermi/GBM so it is sensitive to lensed GRBs with time delays longer than burst duration and shorter than 300 seconds. Here the typical time delay is in order of 10 seconds which corresponds to $M_L=10^6 M_{\odot}$.
\subsection{Simulating Lensing Data}
\label{subsec:Mock}
\begin{figure}[!htbp]
\centering
\includegraphics[width=0.7\linewidth]{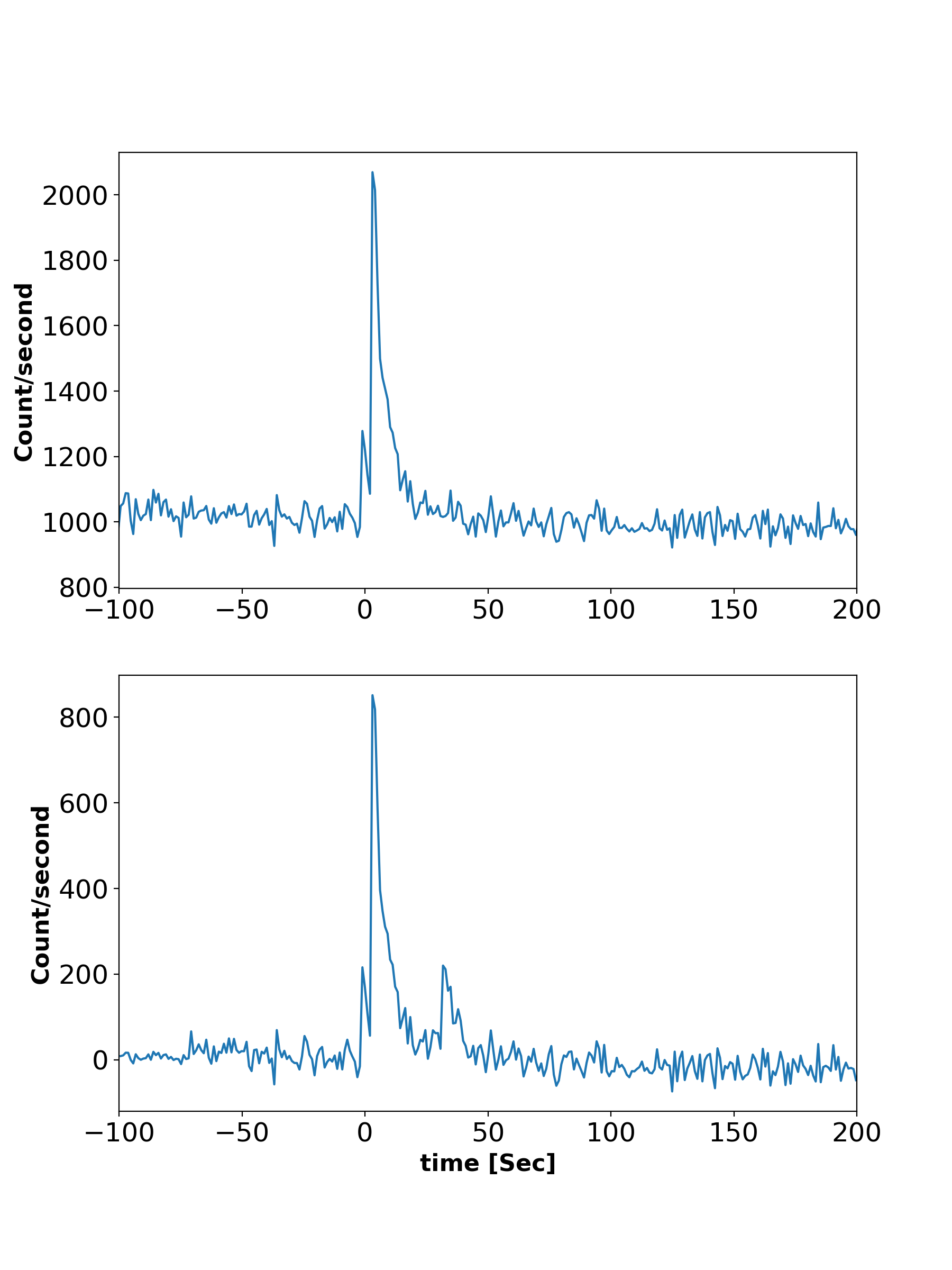}
\caption{\label{fig:mock_Lc} Top: The light curve of an original GRB (GRB 200714774). Bottom: Simulated lensed light curve of the same GRB due to a point lens with a mass of $5 \times 10^5 M_\odot$ at a cosmological redshift of half that of GRB ($z_L=1$ distance $\sim 1Gpc$), and with a normalized impact parameter $\beta/\theta_E = 0.7$. This assumed lens makes two lensed images with a time delay of 28 seconds and a magnification ratio of 4.}
\end{figure}

We simulated a lensed light curve by superimposing the count rate $I(t)$ of a real GRB with itself as shown in Fig.\ref{fig:mock_Lc}, by imposing a time delay $\Delta t$ and a magnification ratio $R$. The resulting lensed light curve count rate is
\begin{equation}
I'(t)=\frac{R}{R+1} I(t) + \frac{1}{R+1}I(t+\Delta t).
\label{eq:superimpose}
\end{equation}
The crucial feature of this equation is that the fainter peak of the light curve, which corresponds to the image with the lower magnification, will arrive later than the brighter one, which corresponds to the image with the higher magnification ({\it i.e.} $t^+ \leq t^-$) as derived from Eqs. (\ref{eq:delay}) and (\ref{eq:magnification}).
\par
For a specific lensing configuration with normalized impact parameter $\beta/\theta_E = 0.7$, lens mass of $5 \times 10^5 M_\odot$ and $D_s=2D_L$ (mentioned in section \ref{sec:GLensing}), and using equations (\ref{eq:delay}) and (\ref{eq:magnification}), we find the corresponding time delay and magnification for the simulated GRB in Fig.\ref{fig:mock_Lc} to be $\Delta t=28 s$ and $R=4$.
\subsection{Autocorrelation Function Technique}
\label{subsec:ACf}
\begin{figure}[!htbp]
\centering
\includegraphics[width=0.7\linewidth]{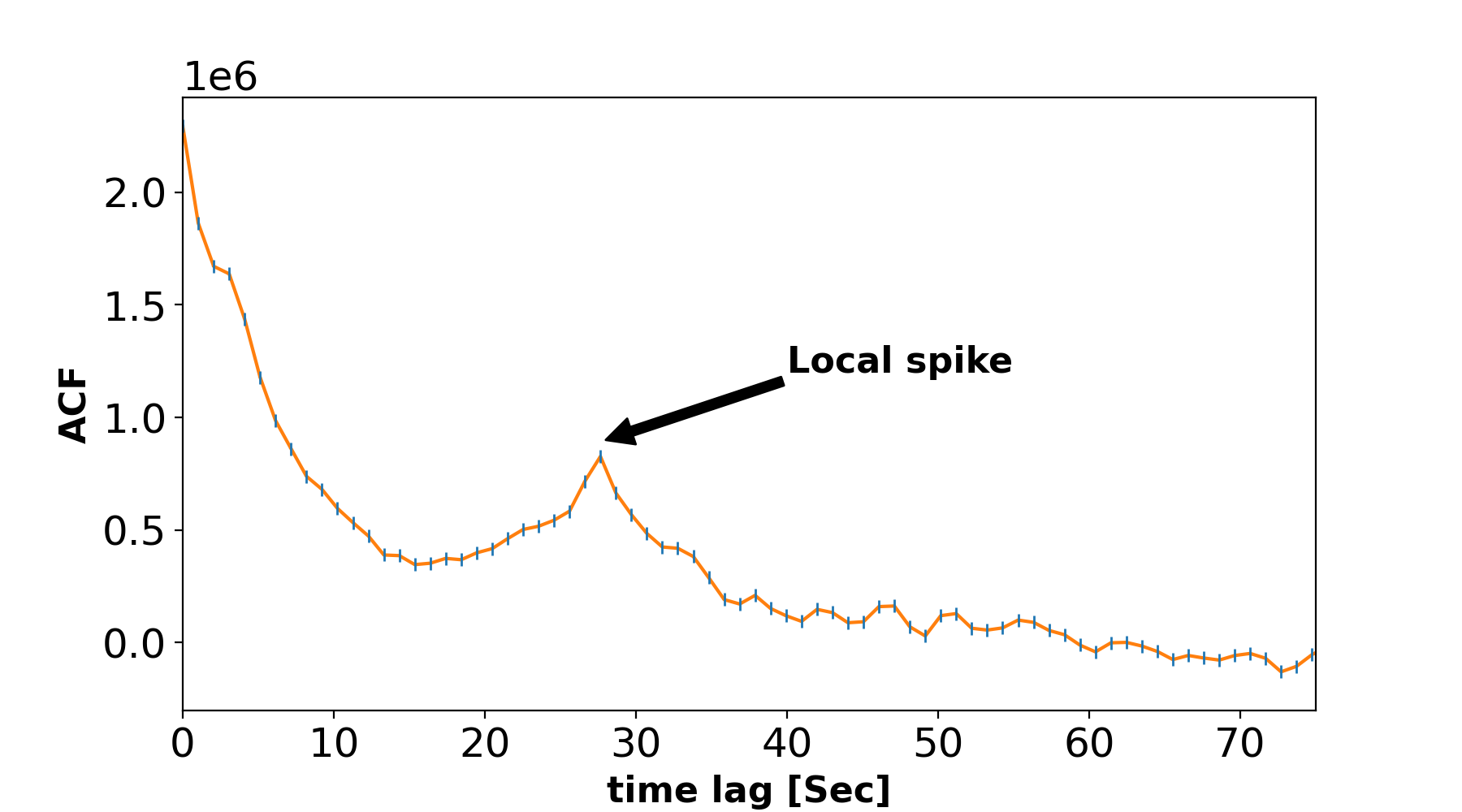}
\caption{Autocorrelation function of the mock lensed light curve of Fig.\ref{fig:mock_Lc} (down) that showing a local peak at the imposed time delay of $28$s.}
\label{fig:mock_ACF}
\end{figure}

We apply the autocorrelation function as a technique to identify two peaks in the GRB light curve due to the two lensed images and to obtain the time delay. The autocorrelation function of count rate $I(t)$ with a time lag of $\tau$ is defined by,
\begin{equation}
C(\tau)=\sum_t^{n(\tau)=T-\tau} I(t)I(t+\tau).
\label{eq:ACf}
\end{equation}
Applying the autocorrelation function on a simulated light curve that contains two local peaks due to two lensed images with an artificial time delay $\Delta t$, will yield a maximum or local spike in $C(\tau)$ at a time lag $\tau$ that corresponds to the delay time $\Delta t$. For the simulated light curve in Fig.\ref{fig:mock_Lc}(bottom panel), a spike is identified at $\tau=28 s$ which represents the time delay between the lensed images as shown in Fig.\ref{fig:mock_ACF}. While \cite{Ji2018StrongLO} used a normalized autocorrelation function to determine the time delay between the gravitationally lensed images, which is less sensitive or yields suppressed spikes, we use a non-normalized autocorrelation function which gives better results and higher  sensitivity.  We note in Appendix \ref{section:appendixA} that the error in estimation of the unnormalized  ($C(\tau)$) and the  normalized ($C_N(\tau)$) autocorrelation function are related together as  $\Delta  C_N(\tau) \geq \sqrt{C(\tau)^2+(\frac{\Delta \Sigma_1}{\Sigma_1})^2+(\frac{\Delta \Sigma_2}{\Sigma_2})^2}$, where $\Sigma_1=\sum I(t)^2 \Delta t$ and $\Sigma_2=\sum I(t+\tau)^2 \Delta t$. This means that unnormalized autocorrelation has smaller error  with respect to the normalized ones.
\par
The standard error (SE) of the autocorrelation function in each lag $\tau$ is defined as 
$SE =\frac {\sigma (\tau) }{\sqrt {n(\tau)}}$,
where $\sigma(\tau)$ is the standard deviation of the population in each lag and $n(\tau)$ is the number of data points in the summation of Eq.(\ref{eq:ACf}). The reported upper and lower limits of the autocorrelation function in each lag are given by the 68\% confidence level as the ${C(\tau)}+( {SE} \times 1.69),$ $ C(\tau)-( {SE} \times 1.69)$, respectively. To estimate the error in $\Delta t$, {\it i.e.} $Err(\Delta t)$, where the autocorrelation function has a maximum at $\Delta t= \tau^*$ (the time delay between the two lensed images), we consider the error corresponding to $\tau^*$ where $\tau^*= \Delta t = \tau^* \pm Err(\Delta t)$. This time intervals assigns the error in the time delay between the two lensed images and if the lag error is less than the time resolution, the time bin is reported as the error in the time delay.
\par
\begin{figure}[htbp]
\centering
\includegraphics[width=0.7\linewidth]{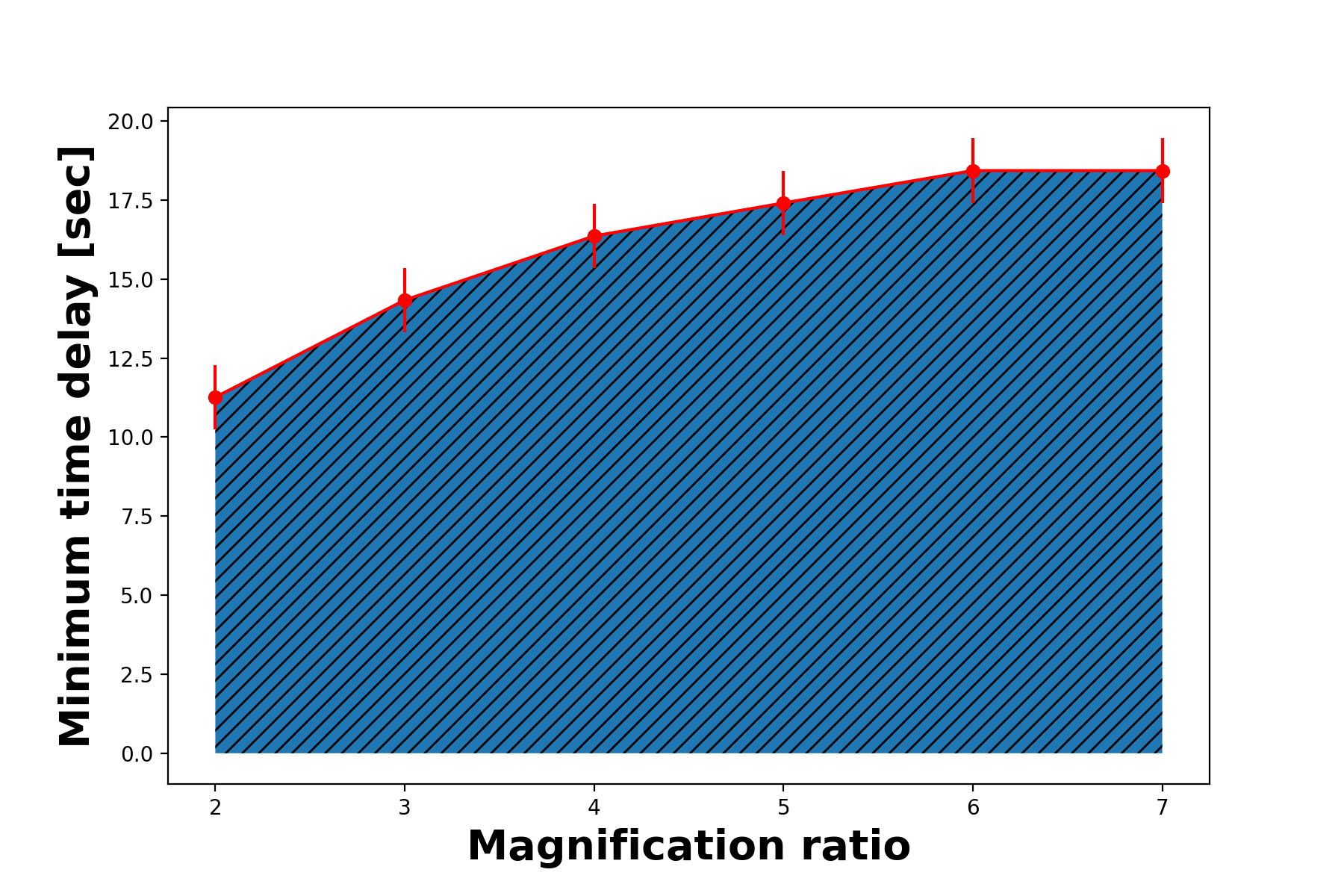}
\caption{An example of the time delay detection sensitivity of the autocorrelation technique as a function of the magnification ratio, applied for the simulated lensed light curve of the burst shown in Fig. \ref{fig:mock_Lc} (top panel)}
\label{fig:minimum_detecable}
\end{figure}

For the burst is shown in Fig.\ref{fig:mock_Lc} (top panel), we generate mock lensed GRB with an imposed magnification ratio, then decrease time delay between two gravitational lensed images until the autocorrelation technique can not detect the time delay between these two images. In Fig.\ref{fig:minimum_detecable} the excluded part of figure (shaded) is the time delay below the sensitivity of autocorrelation technique in each magnification ratio. In this example, we see that a decrease in the magnification ratio of the two light curve local peaks makes it more sensitive to detect shorter time delays. The minimum detectable time delay for a large magnification ratio is about the width of the peak. As a result, this method is not sensitive to a delay that is less than the width of each peak in the light curve ({\it i.e.} peak duration).

\subsection{Magnification Ratio}
\label{subsec:R}
To find the magnification ratio of real lensed GRB light curves, we subtract the background count rate. The magnification ratio can be sensitive to the background model. Background intervals were selected before and after the burst at about the same duration of the burst interval and with little overlap with the burst interval, then the lowest order polynomial is chosen to fit the background.
\par
For the simulated light curve, we set the polynomial order to be zero (a constant). We first remove the background from the simulated count rate and then the superimposed light curves according to equation (\ref{eq:superimpose}) (Fig.\ref{fig:mock_Lc} bottom panel). Retrieved magnification ratio calculated by the ratio of the two local peaks of this simulated lensed GRB is $R=3.442\pm 0.512$, which approximately equals the imposed magnification ratio (R = 4).

\subsection{Instrumental Error}
\label{subsec:instrumental-error}
Due to purely random distribution of count rate error (which is reported in ASCII output of RMFIT tool), for each GRB candidate, we generated one hundred Gaussian realization of its count rate. Then we used the autocorrelation technique of each realization to find the time delay and the lag error in the time delay. On the other hand, we computed the magnification ratio and propagated the error in the magnification ratio for the individual realizations. Finally, we calculate the mean of these parameters and report them as the time delay and magnification ratio with the corresponding $1-\sigma$ uncertainty.

\subsection{Efficiency and Accuracy of the Detection Technique}
\label{sec:Efficiency}
\begin{figure}[htbp]
\centering
\includegraphics[width=0.7\linewidth]{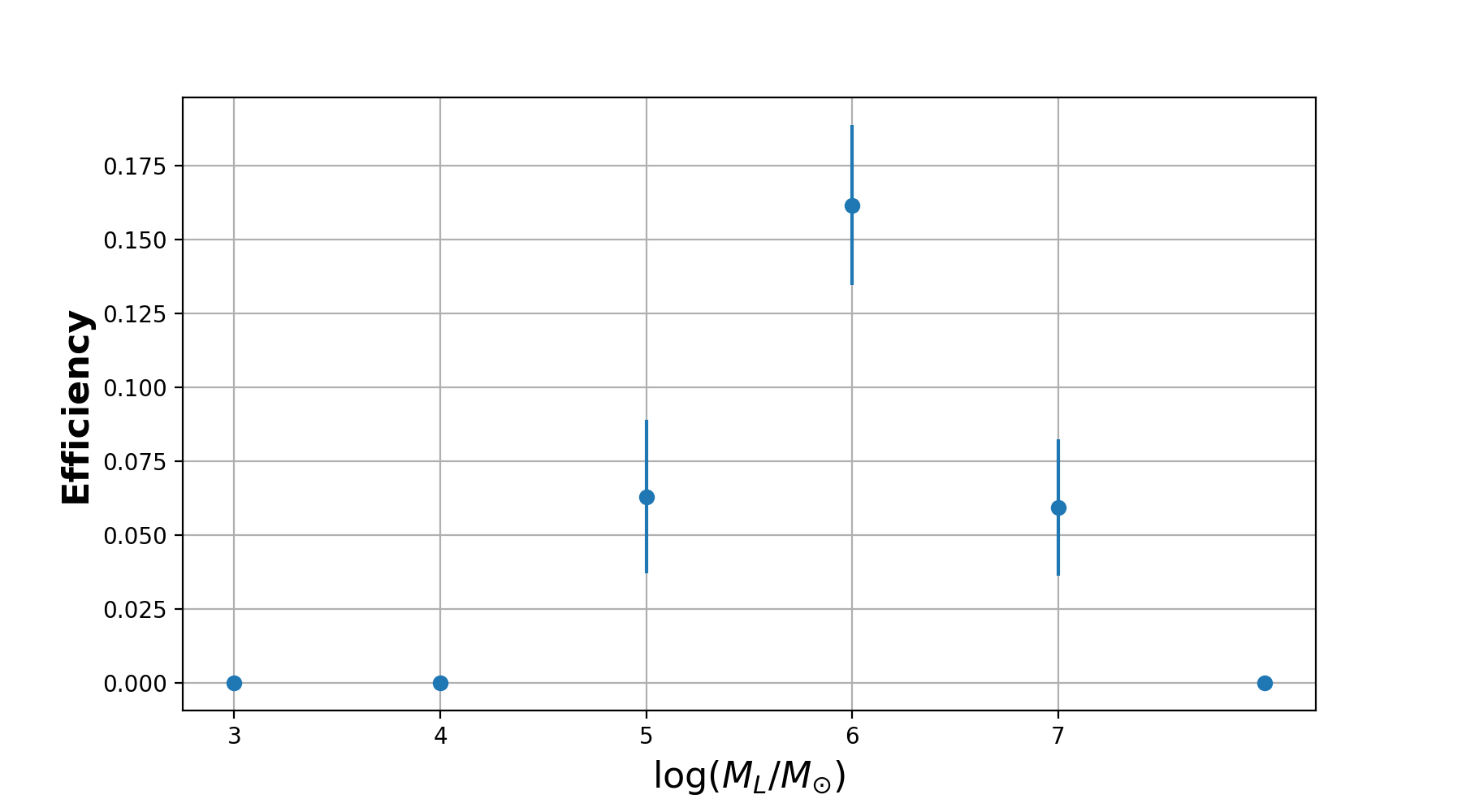}
\caption{The efficiency of the autocorrelation technique versus the lens mass in a logarithmic scale as obtained from the Monte-Carlo simulation of  gravitational lensed GRBs}
\label{fig:efficiency}
\end{figure}

We performed a Monte-Carlo simulation of mock lensed GRBs to calculate the efficiency of the Autocorrelation function technique in finding lensed GRBs. In this simulation, we first selected 81 long GRBs from the Fermi/GBM catalog which just have one peak in their light curves, so they have no local peaks in their natural autocorrelation function. Because of using TTE data files of GRBs light curves which maximally records 300 seconds after the burst, and due to the set time bin of 64 milliseconds, gravitational time delay less than 64 milliseconds and greater than 300 seconds that correspond to the lens mass of $M_L \leq 10^2 M_{\odot} $ and $M_L \geq 10^9 M_{\odot} $ are excluded from our simulation. So we simulated lensed GRBs with a lens mass in the range of $10^3-10^8 M_\odot$. We assumed a point mass lens and defined mass bin of $3 \leq  i=\log (M_L/M_{\odot})\leq 8$ then in each mass bin of $i$ we generated 810 mock lensed data ($N_{Generated}=810$) with random parameters of normalized impact factor $y=\beta/\theta_E$ and source and lens cosmological redshifts ($z_s\leq 7$ and $z_L\leq z_s$) then we applied the autocorrelation method for these simulated GRBs to evaluate efficiency in detecting of the imposed time delay by autocorrelation technique. Also we checked that the retrieved magnification ratio of the simulated light curve is within the confidence interval of $2 \sigma_R$ ($\bar{R}-2\sigma_R \leq R \leq \bar{R}+2\sigma_R$). If for a simulated lensed GRB we find the conditions concerning the detection of 
the imposed time delay and magnification ratio are fulfilled, we consider it a detected event. We defined the efficiency of our detection method in $i^{th}$ mass bin as
\begin{equation}
\epsilon _i= \frac{N_{i}}{N_{Generated}},
\label{eq:epsilon}
\end{equation}
where $N_i$ is detected lensed GRB in each mass bin of $i$. The result of the simulation is depicted in Fig.\ref{fig:efficiency} that shows the autocorrelation technique has the best performance in the mass about of $10^6 M_\odot$ for the gravitational lenses. 
\par
Efficiency parameter must be used in the calculation of optical depth by normalizing the number of lensed candidates
\begin{equation}
N_{Detected}= \sum_i \frac{N_i}{\epsilon _i}.
\label{eq:Ncorrected}
\end{equation}
So the optical depth in Eq.(\ref{eq:tauTotal}) should be rewritten as 
\begin{equation}
\tau_{obs}=\frac{1}{N_{Total}} \sum_i \frac{N_i}{\epsilon _i},
\label{eq:tauObserved}
\end{equation}
As we will shown in the section \ref{sec:Results}, we find one candidate of lensed GRB with mass in order of $10^6 M_\odot$, therefore the numerical value of $\tau_{obs}$ should be $\tau_{obs}=\tau_{Average}/\epsilon_6= (1.340\pm 0.213) f$. We
use this result to find the contribution of gravitational lenses with mass in order of $ 10^6 M_{\odot}$ to the matter content of Universe in the section \ref{sec:Results}.
\par
\begin{figure}[htbp]
\centering
\includegraphics[width=0.7\linewidth]{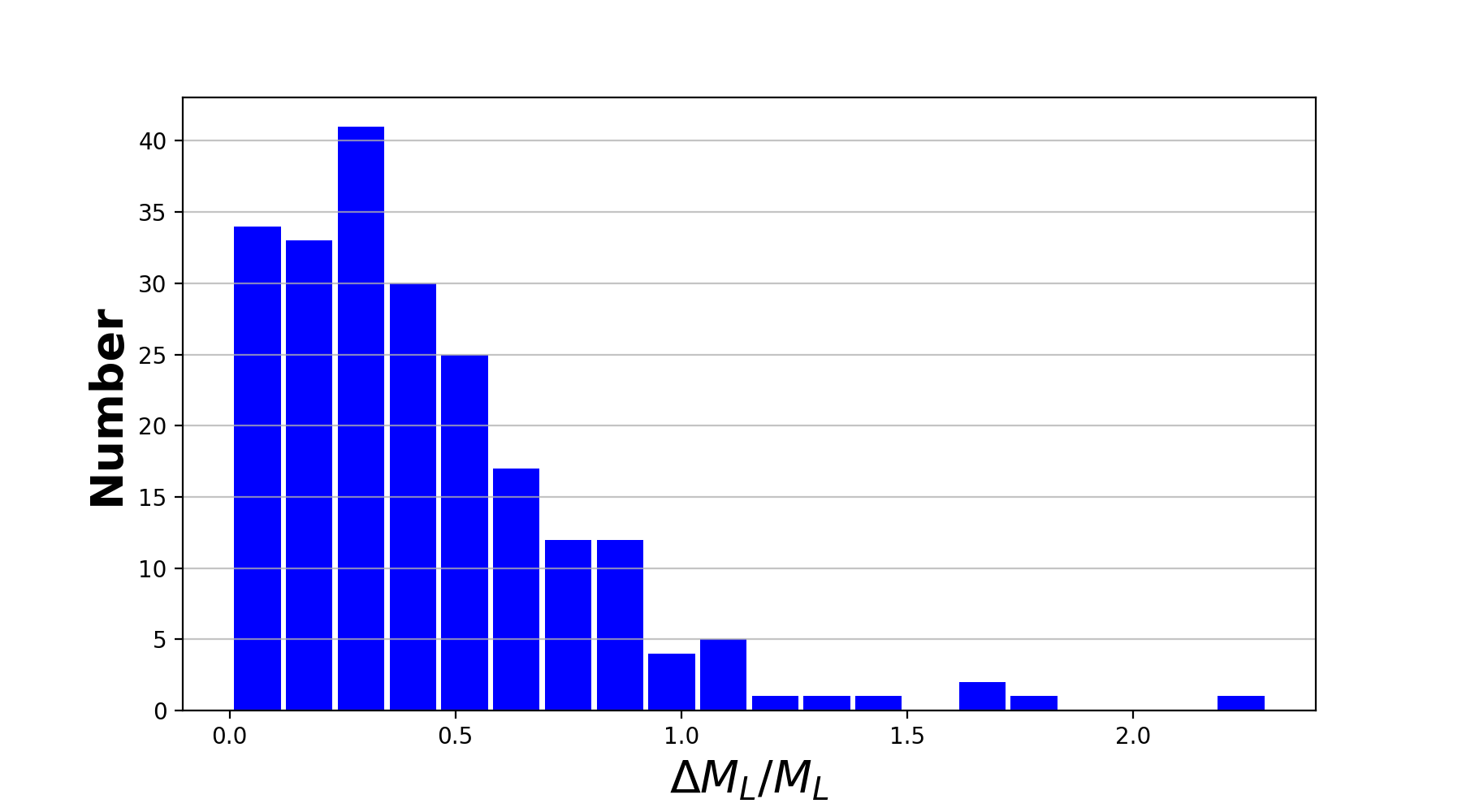}
\caption{Histogram of $\Delta M_L/M_L$ for detected events from the simulated lensed GRBs. In our method of detection, 94.54\% of the detected simulated lensed GRBs have $\Delta M_L/M_L <1$}
\label{fig:accuracy}
\end{figure}

To find the accuracy of our method, for detected event in our simulation, we retrieve the magnification ratio and the time delay from the lensed GRB light curve, then we use equations (\ref{eq:delay}) and (\ref{eq:magnification}) to retrieve the  mass lens. The histogram of the difference between the determined mass from the simulated light curve and the imposed lens mass to the imposed lens mass($\Delta M_L/M_L$) is depicted in Fig.\ref{fig:accuracy} that shows 94.54\% of detected simulated lensed GRBs with our method have $\Delta M_L/M_L<1$. We note that uncertainty in mass implies an uncertainty to the equation \ref{eq:tauObserved} (\emph{i.e} $\epsilon_i (M_L)$). Since the efficiency function in Fig. (\ref{fig:efficiency}) varies in the logarithmic scale of lens mass, we can ignore our mentioned uncertainty of mass in the optical depth calculation.

\section{DATA ANALYSIS}
\label{sec:DataAnalysis}
\subsection{Catalog}
\label{subsec:Catalog}
To search for lensed GRB candidates, we considered long GRBs then ignored the faint GRBs with fluence less than $10^{-7}\ erg/cm^2$ and the GRBs whose light curves were not recorded completely. As a result, out of 2965 GRBs recorded during 2008-2020 by Fermi/GBM, we found 2137 GRBs that matched these criteria. We investigated all 2137 GRBs in this sample and applied the search technique that is mentioned in the next subsection.

\subsection{Selection Criteria}
\label{subsec:Criteria}
\begin{itemize}
    \item 
    We select the GRB candidates that show two distinguishable peaks in their light curves with the faint peak following the bright peak in the energy range over which the GRB spectrum did not show overflow or cut-off effects that usually do not happen between 8 KeV-150 KeV.
    \item We chose the NaI detector that shows the brightest signal and for the avoidance of instrumental error, we also checked that at least three other bright NaI detectors recorded the two peaks for the selected GRB in the different energy ranges. This is to ensure a better signal-to-noise ratio and good statistics. Ten candidates of the 2137 GRBs passed these two criteria.
    \item Our autocorrelation technique requires the time difference between two peaks of the light curve to be greater than the full width of one peak to show extremum near the time delay between the two peaks.
\end{itemize}
\par
\begin{figure}[htbp]
\centering
\includegraphics[width=0.7\linewidth]{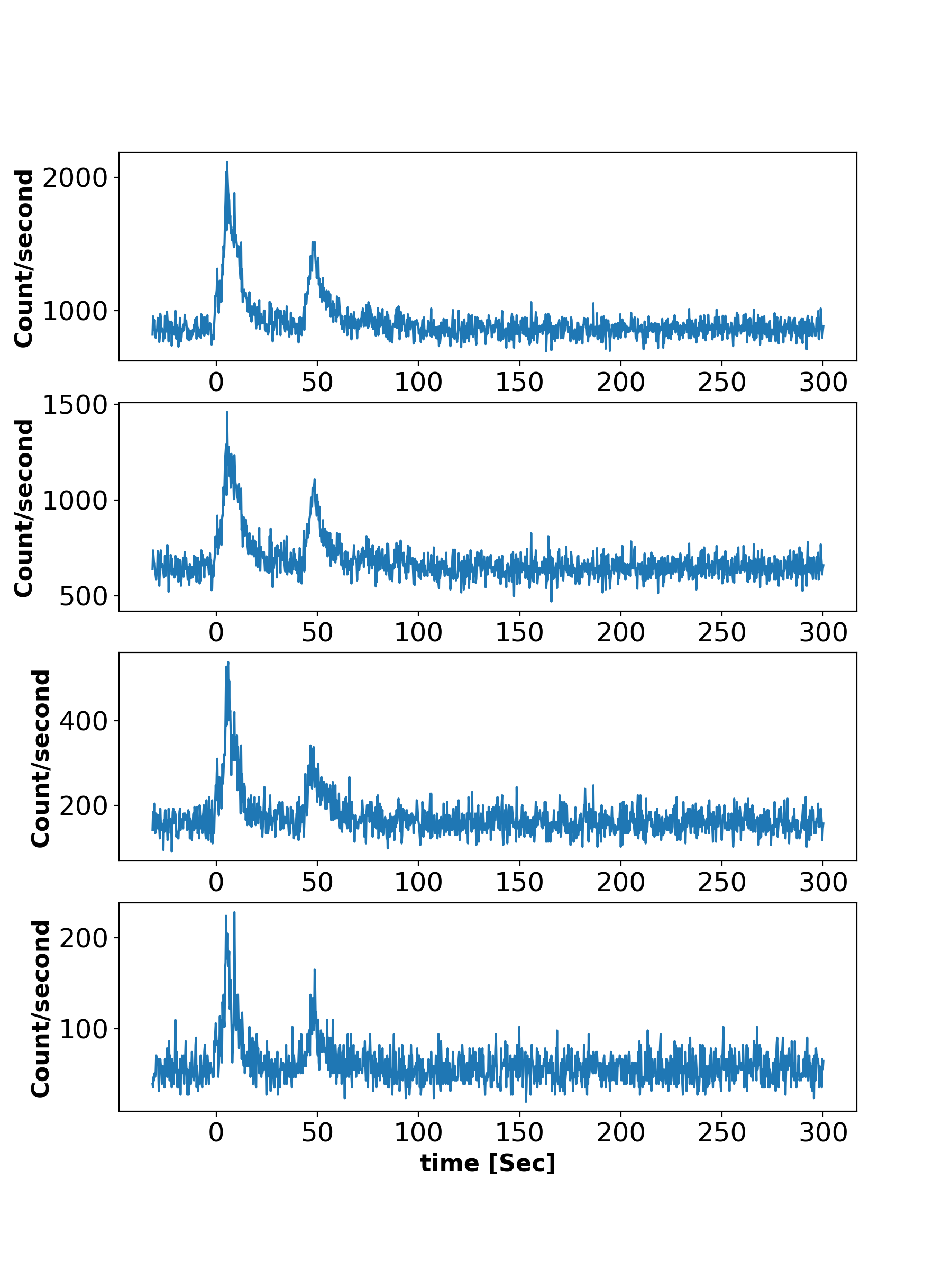}
\caption{Light curve of the gravitationally lensed GRB candidate: GRB 090717034. The energy range of the panels are 8-150 keV, 8-50 keV, 50-100 keV, 100-150 keV from top to bottom.}
\label{fig:090717034_LC}
\end{figure}
\begin{figure*}[!htbp]
\begin{minipage}{0.47\textwidth}
\includegraphics[width=\linewidth]{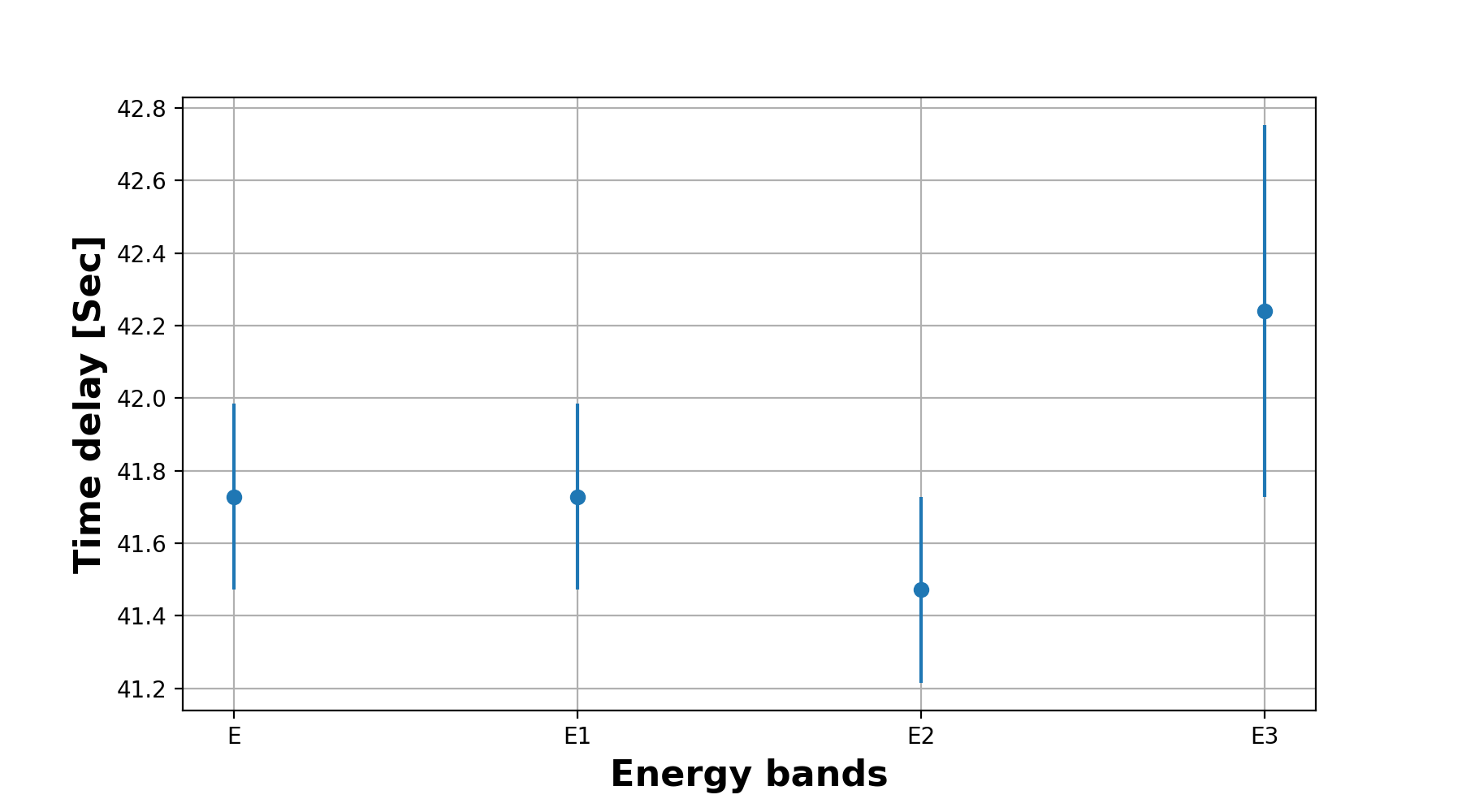}
\end{minipage}
\hfill
\begin{minipage}{0.47\textwidth}
\includegraphics[width=\linewidth]{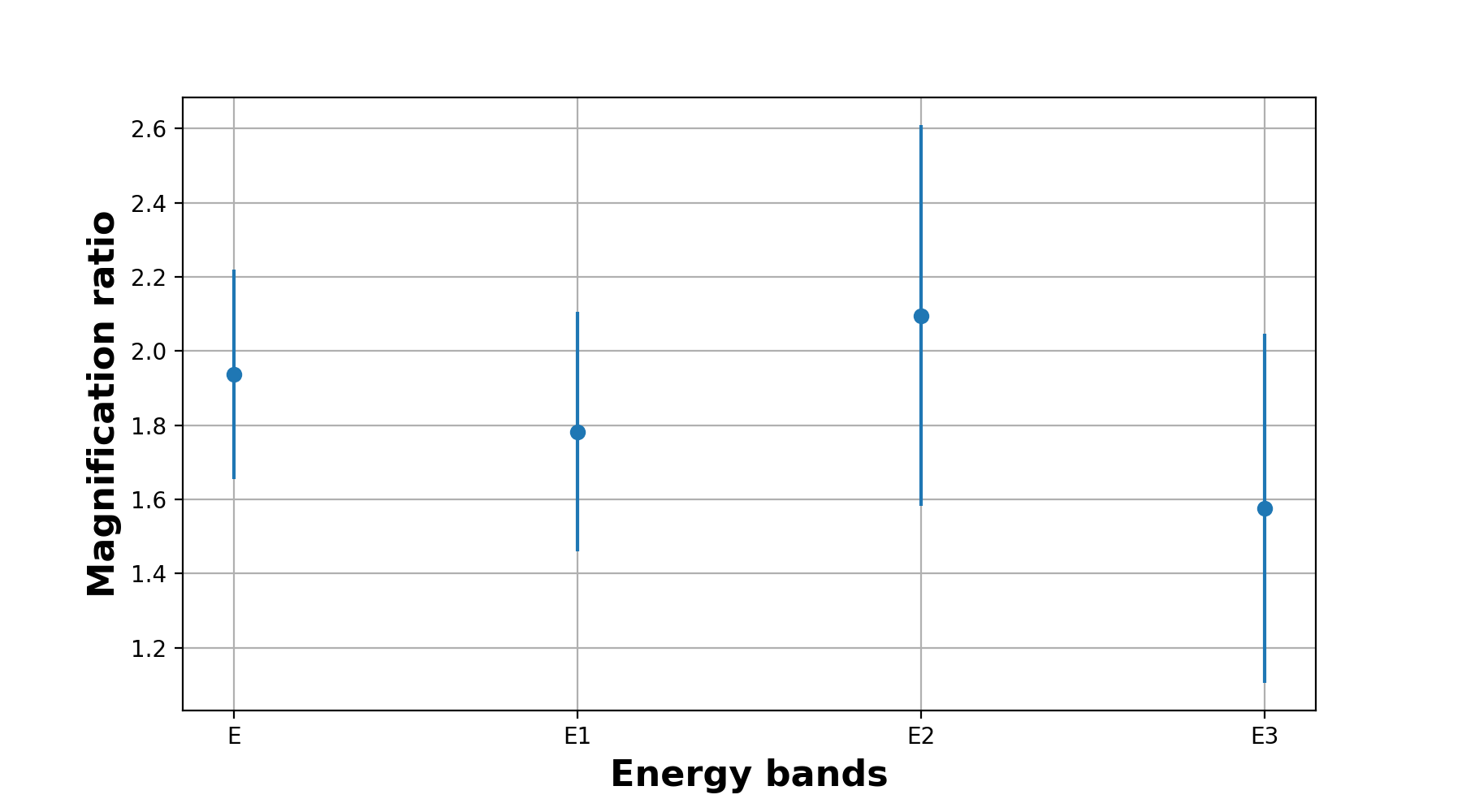}
\end{minipage}

\caption{Time delay (Left) and magnification ratio (Right) of the gravitationally lensed candidate GRB 090717034 in the Fermi GBM full energy range E (8-150 keV) and in three energy bands E1 (8-50 keV), E2 (50-100 keV) and E3 (100-150 keV) with the one sigma confidence level shown. The time delays and magnification ratios in four bands of energy depicts consistency at the level of one sigma that verified this GRB can be regarded as a gravitational lensed candidate.}
\label{fig:090717034}
\end{figure*}
\begin{figure}[htbp]
\centering
\includegraphics[width=0.7\linewidth]{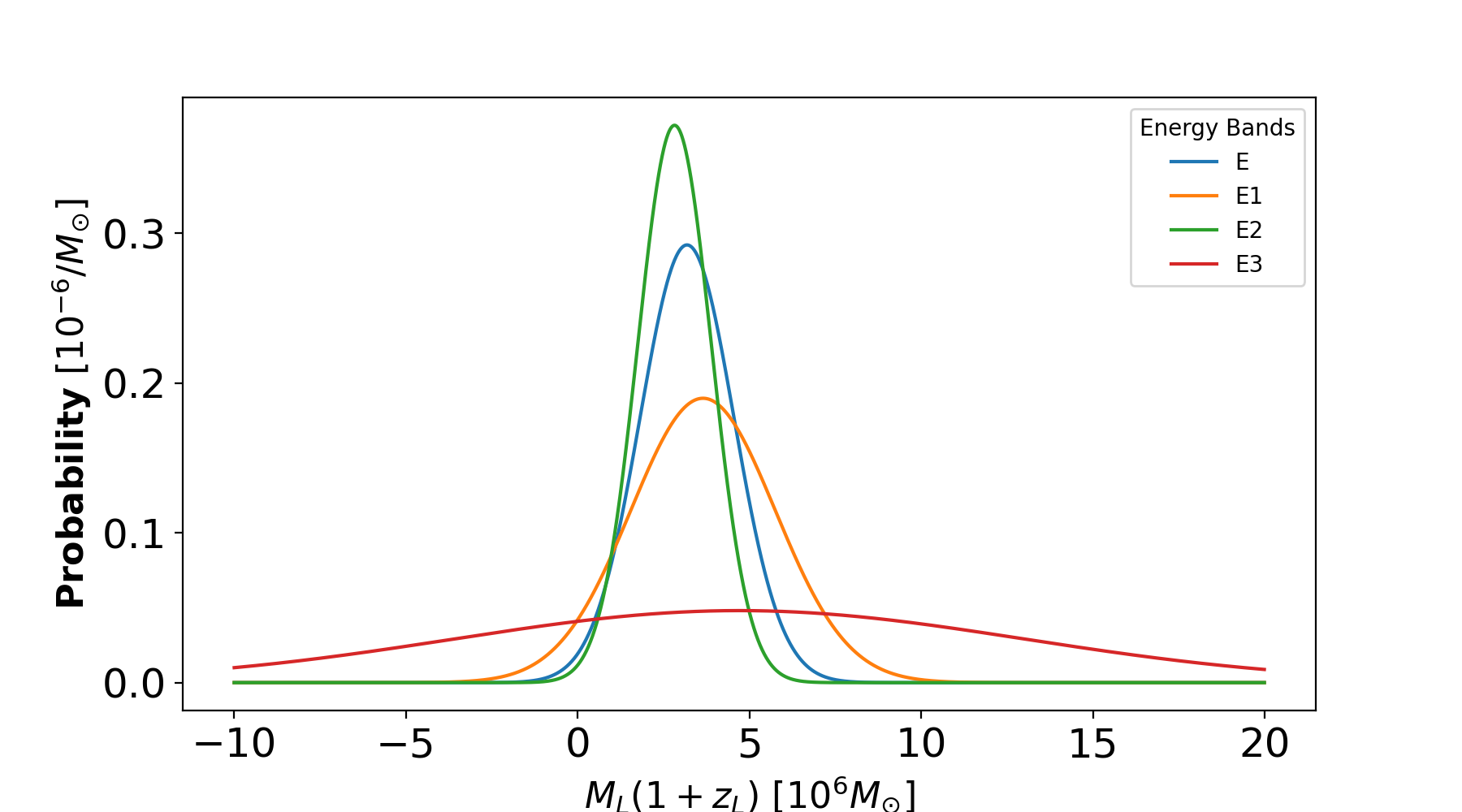}
\caption{Probability distribution of the lens mass. The labels of E, E1, E2, and E3 correspond to the four energy bands in the ranges of 8-150 keV, 8-50 keV, 50-100 keV, 100-150 keV. The coverage of probability distribution in four energy ranges manifests the estimate redshifted mass is independent of energy of photons.}
\label{fig:mass}
\end{figure}

Since the time delay and the magnification ratio of the two lensed images are independent of the photon energy (Eqs. (\ref{eq:delay}) and (\ref{eq:magnification})), we anticipate that these parameters will remain the same in the different energy channels. By defining three energy intervals between 8 keV to 150 keV (8-50 keV, 50-100 keV, 100-150 keV),we should check that GRBs light curves have two peaks and their flux ratios and the time delays are similar (at the level of 1-$\sigma$) in each energy interval. This requirement assures that the two peaks of the GRB candidates are real and not the result of an artifact. GRBs that manifested all these features can be regarded as candidates for gravitationally millilensed GRBs. We find one candidate that fulfills each of the above criteria (GRB 090717034) and its light curve is shown in Fig.\ref{fig:090717034_LC}. For each the light curve in the full energy range and the three ranges depicted in Fig. \ref{fig:090717034_LC}, we derive the time delay and magnification ratio with their one sigma error. The result is depicted in Fig.\ref{fig:090717034} that shows both the magnification ratio and time delay in the four energy channels show a consistent behavior. For this verified millilensed candidate, substituting the time delays and magnification ratios in each of four energy channels (8-50 keV, 50-100 keV, 100-150 keV and 8-150 keV) in equations (\ref{eq:delay}) and (\ref{eq:magnification}), we determine the redshifted masses of the lenses $M_L(1+z_L)$ in the three energy bands and the full one. The probability of mass is depicted in Fig.\ref{fig:mass} that shows consistency of the estimated redshifted massed in the different energy bands.

\begin{table*}[!htbp]

\caption{\bf Estimated Redshifted Mass}
\begin{tabular}{ccccc}
GRB ID &$\Delta t$ &  R & $M_L(1+z_L)$ \\
\hline
GRB 090717034 & $41.332\pm 0.283$ s & $1.782\pm 0.239$ & $(3.615\pm 1.763)\times10^{6}M_\odot$\\

\end{tabular}
\label{tab:mass}
\end{table*}


\section{RESULTS AND CONCLUSION}
\label{sec:Results}
We presented an investigation and a technique to probe gravitational lensing in GRB due to a lensing configuration that can be practically assessed with the Fermi observations through autocorrelation of the echo signals in the GRBs light curves. We first screened the 2008-2020 GRBs and excluded short GRBs and faint bursts (fluence $\leq 10^{-7}  erg/cm^2$) and those whose light curves were not recorded completely. This reduced the sample from 2965 to 2137 bursts. We found one candidate of lensed GRBs that satisfied all criteria of section \ref{sec:DataAnalysis}. In order to calculate the uncertainty in the time delay and magnification ratio for this gravitational lensed candidate, we generated one hundred Gaussian realizations from the time series of this candidate count rate, then obtained its mean time delay and magnification ratio with the corresponding errors in the overall energy band (8-150 keV), and then estimated the redshifted mass of the lenses $M_L(1+z_L)$. The result is reported in Table \ref{tab:mass} represents the lens redshifted mass in the order of $\approx 10^{6} M_\odot$, and since we assumed a point mass model for the lens, this compact lens can be regarded as a supermassive black holes (SMBHs). Primordial Black Holes (PBHs) with a mass of $10^5  M_\odot$ may provide seeds for SMBHs and different authors have been proposed diverse mechanisms to explain the formation of supermassive PBHs (\cite{Mehrab13,Mehrab14,Mehrab15,Mehrab16}). A plausible scenario is that PBHs had formed from large-scale peaks in density fluctuations in  the early universe (\cite{Mehrab12}).
\par
To find the expected mass fraction of the compact lenses with mass in the order of $10^{6} M_\odot$ to the total matter mass, we used Eqs. (\ref{eq:tauSSimplified}) and (\ref{eq:tauObserved}), which give an observed optical depth equals to the ratio of the number of lensed candidates to the total number of GRB in our sample of GRBs, normalized by the detection efficiency in the  $10^6 M_{\odot}$ mass bin obtained from our Monte-Carlo simulation. We find one millilensed candidate out of 2137 GRBs, therefore, $\tau_{Average}=1/2137$ and with considering the Poisson uncertainty, the optical depth will be $\tau_{Average}=(2.033\pm 1.115)/2137$. So the observed optical depth can be found as $\tau_{obs}=\tau_{Average}/\epsilon_{6}=(2.033\pm 1.115)/(2137*\epsilon_{6})$. Therefore, the density parameter of these black holes (For the Dirac mass function with masses $\approx 10^6 M_{\odot}$ ) would be $\Omega_{BH} \approx 0.007 \pm 0.004$ so the fraction of BHs to the matter $\Omega_{BH}/\Omega_M=0.027 \pm 0.016$ (Eq.(\ref{eq:fDM})). Our result is one order of magnitude large compared to previous work of \cite{Population3} in the mass range of $10^2 -10^3 M_{\odot}$ where they found one lensed candidate out of 1821 GRBs examined, so the average optical depth could be determined from $\tau_{Average}=1/1821=5.49 \times10 ^{-4}$ and as a result, the density parameter in this mass range would be $\Omega_{BH} \approx 5 \times 10^{-4}$. Recently, \cite{2021NatAs} probed the mass range of $10^2-10^7 M_{\odot}$ and reported $\Omega_{BH} \approx 4.6 \times 10^{-4}$ for the point mass model of gravitational lens.
\par
 GRBs as the most distant point-like sources are ideal sources for probing the Universe. Gravitational lensing of these sources can identify the compact and non-compact structures in the Universe. In order to have better observations of millilensing of GRBs, more sensitive detectors with higher resolution to better discern the GRB light 
curves will enable us to find more lensing events and hence understand the matter content of the Universe.


\section*{ACKNOWLEDGEMENT}
This work was supported by SQU grant IG/SCI/PHYS/20/03.
\par

\appendix
\section{ COMPERING ERROR OF AUTOCORRELATION AND NORMALIZED AUTOCORRELATION}

\label{section:appendixA}
The normalized autocorrelation function is defined as (\cite{Ji2018StrongLO})
\begin{equation}
    C_N(\tau)=\frac{<I(t) I(t+\tau)>}{\Sigma_1 \Sigma_2}=\frac{C(\tau)}{\Sigma_1 \Sigma_2},
    \label{eq:normalizeACF}
\end{equation}
where $C(\tau)$ is autocorrelation function and we defined  $\Sigma_1=\sum I(t)^2 \Delta t$ and $\Sigma_1=\sum I(t+\tau)^2 \Delta t$. To compare the error of normalized autocorrelation and autocorrelation, we first take the logarithm of both sides of Eq.\ref{eq:normalizeACF}
\begin{equation}
    \ln  (C_N(\tau))=\ln (C(\tau))-\ln(\Sigma_1) -\ln( \Sigma_2),
    \label{eq:ln_normalizeACF}
\end{equation}
taking derivative of both sides concludes
\begin{equation}
   \frac{\Delta  C_N(\tau)}{ C_N(\tau)}=\frac{\Delta  C(\tau)}{ C(\tau)}-\frac{\Delta \Sigma_1}{\Sigma_1}-\frac{\Delta \Sigma_2}{\Sigma_2}
    \label{eq:DeltaNormalized}
\end{equation}
therefore,
\begin{equation}
 \Delta  C_N(\tau) \geq \sqrt{C(\tau)^2+(\frac{\Delta \Sigma_1}{\Sigma_1})^2+(\frac{\Delta \Sigma_2}{\Sigma_2})^2}
    \label{eq:CompareDelta}
\end{equation}
As a result normalized autocorrelation error is greater than the error of autocorrelation function.
\par
To find autocorrelation error, first we write the autocorrelation function as
\begin{equation}
    C(\tau)=<I(t) I(t+\tau)> =\ \sum_{I_1}\sum_{I_2} I_1 I_2 \ p(I_1, t; I_2, t+\tau) \Delta I_1 \Delta I_2,
    \label{eq:ACF}
\end{equation}
where  $p(I_1, t; I_2, t+\tau)$ is the two point joint probability density function. To find error of of autocorrlation function we take derivative of Eq.\ref{eq:ACF}
\begin{equation}
    \Delta C(\tau) = \sum_{I_1}\sum_{I_2} I_1 I_2 \ \Delta p(I_1, t; I_2, t+\tau) \Delta I_1 \Delta I_2,
    \label{eq:DeltaACF}
\end{equation}
the joint probability density function is defied from joint probability as 
\begin{equation}
  p(I_1, I_2)=\frac{P(I_1, I_2)}{ \Delta I_1 \Delta I_2}
    \label{eq:probDensity}
\end{equation}
where joint probability $P(I_1, I_2)$ is
\begin{equation}
  P(I_1, I_2)=\frac{n(I_1, I_2)}{N}
    \label{eq:prob}
\end{equation}
here ${n(I_1, I_2)}$ is the number of events in the bin centered around $(I_1,I_2)$ with width  $\Delta I_1 \Delta I_2$. 
So the derivative of Eq.\ref{eq:probDensity} can be write as
\begin{equation}
   \Delta p(I_1, I_2)=\frac{\Delta P(I_1, I_2)}{ \Delta I_1 \Delta I_2}=\frac{\Delta n(I_1, I_2)}{N\ \Delta I_1 \Delta I_2}
    \label{eq:DelatprobDensity}
\end{equation}
$n(I_1, I_2)$ follows the binomial distribution so that
\begin{equation}
   \Delta n^2(I_1, I_2)=N \ P(I_1, I_2) (1-P(I_1, I_2))
    \label{eq:Delta_n}
\end{equation}
substitution Eq.\ref{eq:Delta_n} in Eq.\ref{eq:DelatprobDensity} results
\begin{equation}
   \Delta p(I_1, I_2)=\frac{\sqrt{N \ P(I_1, I_2) (1-P(I_1, I_2))}}{N\ \Delta I_1 \Delta I_2}
    \label{eq:DelatprobDensity1}
\end{equation}
For $P(I_1, I_2)\ll 1$, we can estimate above equation as
\begin{equation}
   \Delta p(I_1, I_2)  \approx \frac{\sqrt{N \ P(I_1, I_2)}}{N\ \Delta I_1 \Delta I_2}.
    \label{eq:DelatprobDensity1}
\end{equation}
Therefore, the variance of joint probability function can be found as
\begin{equation}
   \sigma (p(I_1, I_2))=\frac{\Delta p(I_1, I_2)}{\sqrt{n(I_1, I_2)}}\approx \frac{\sqrt{ \ P(I_1, I_2)}}{ \Delta I_1 \Delta I_2 \sqrt{N\ n(I_1, I_2)}}.
    \label{eq:DelatprobDensity1}
\end{equation}
So $\sigma (p(I_1, I_2))$ is the error of joint probability density.

\newpage
\bibliographystyle{unsrtnat}
\bibliography{References}


\end{document}